\documentclass[preprint,nofootinbib,superscriptaddress,amsmath,amssymb]{revtex4}

\usepackage{amsmath} 
\usepackage{verbatim} 
\usepackage{graphicx} 

\begin{document}

\title{On the Scale of Inflation}

\author{C. Armendariz-Picon}
\affiliation{Department of Physics, Syracuse University, Syracuse, NY 13244-1130, USA}
\email{carmenda@syr.edu}

\begin{abstract}
Because the scale of inflation  is conformal frame dependent,  in order to fully characterize it   one should quote its value in terms of all the independent  equal-time  dimensionless ratios in the theory. We argue that when  couplings  depend on the inflaton itself,  one cannot calculate these  ratios in  terms of measurable quantities such as the tensor amplitude and the Planck mass at present.   This uncertainty  also becomes manifest  when we try to express the (frame-dependent) inflationary scale in terms of masses that are measurable today.  Although we can calculate inflationary scale in the Einstein frame  in terms of  today's Planck mass, we cannot do the same in the Jordan frame. There are thus grounds to claim  that the tensor amplitude does not completely characterize the scale of inflation.

\end{abstract}
\maketitle

\section{Introduction}
Conventional inflationary models predict a specific and robust \cite{Ozsoy:2014sba} relation between the appropriately normalized root mean square amplitude of the primordial tensor modes $A_T$, the scale  of inflation $H_I$ and the Planck mass $M_P$ \cite{Mukhanov:1990me},
\begin{equation}\label{eq:scale}
	H_I= A_T M_P.
\end{equation}

Therefore, it is often argued that a detection of  primordial tensor perturbations  would  reveal the energy scale at which inflation took place, thus justifying considerable  experimental efforts  to achieve such a detection  \cite{Baumann:2008aq}. In this brief note we point out that in general scalar-tensor theories, the ``scale of inflation"  is an ambiguous quantity, because it is conformal frame dependent. If couplings do not depend on the inflaton, this ambiguity has no consequence when the scale of inflation is calculated using equation (\ref{eq:scale}).   But in theories in which the inflaton does determine the couplings of the theory,  equation (\ref{eq:scale}) may return the wrong inflationary scale when calculated  using the measured value of $M_P$ today. 

In the Einstein frame, inflaton-dependent  couplings are generically  expected to appear in any theory of inflation.  This  is what happens for instance in a quite general but still somewhat restricted class of  scalar-tensor theories similar to those originally proposed by  Jordan,  Brans and Dicke      \cite{Jordan:1959eg,Brans:1961sx}. This class has been widely studied in the cosmological literature \cite{Faraoni:2004pi}, and, in fact, the first inflationary model belongs to this set \cite{Starobinsky:1980te, Mukhanov:1981xt}.  Within this framework, equation (\ref{eq:scale}) returns the correct inflationary scale in the Einstein frame, in which the Planck mass is a constant. But in the Jordan frame, where the proton mass    remains constant,   not only does equation (\ref{eq:scale}) return the wrong inflationary scale under the same assumptions, but  it also remains  practically impossible to determine it in terms of presently observable quantities. Because the choice of conformal frame is simply a matter of convenience, one could hence  argue that in the class of theories considered here a measurement of the tensor amplitude does not fully characterize the scale of inflation. Further consequences of related ideas are  explored for instance in \cite{Ashoorioon:2014yua,Piao:2011bz}.

\section{Scalar Tensor Theories}

To illustrate the limits of  equation (\ref{eq:scale}), let us consider the  relatively general class of scalar tensor theories alluded to in the introduction \cite{Jordan:1959eg,Brans:1961sx}. To formulate these theories, one needs to pick  a conformal frame first. This choice is just a matter of convenience, somewhat analogous to a choice of gauge, and there is no particular ``correct" choice. Physical predictions remain of course conformal frame independent \cite{Chiba:2013mha}.

In order to quantize cosmological perturbations, for instance,  it is more convenient to formulate these theories in the Einstein frame, in which both the kinetic terms for the metric $g_{\mu\nu}$ and the inflaton $\varphi$  assume  canonical forms.  After an inflaton-dependent  conformal transformation and a field redefinition the action of these theories takes the form
\begin{equation}\label{eq:S E}
S_E=\int d^4x \sqrt{-g}\left[\frac{M^2}{2}R-\frac{1}{2}\partial_\mu\varphi \partial^\mu\varphi-V(\varphi)\right]+S_m[F^2(\varphi)g_{\mu\nu},\psi],
\end{equation}
which we shall take to be the starting point of our considerations. For illustration we assume that matter consists of a single scalar $\psi$  of mass $m$, which, by abuse of simplicity we shall call ``the proton",
\begin{equation}
	S_m[g_{\mu\nu},\psi]\equiv \int d^4x \sqrt{-g} \left[-\frac{1}{2}\partial_\mu \psi \partial_\mu\psi-\frac{1}{2}m^2\psi^2\right].
\end{equation}
 The important point here is that the conformal transformation  to the Einstein frame renders the mass parameter $M$ constant, and thereby introduces  a coupling between the matter fields $\psi$ and the inflaton $\varphi$ through the (dimensionless) function $F(\varphi)$, which we take to be arbitrary. Although this class of theories is not particularly natural \cite{ArmendarizPicon:2011ys}, it provides a simple realization of the scenario we want to describe. We expect the same ideas to apply in more realistic theories, in which the tree-level couplings of the inflaton to matter are  not universal. Such non-universal couplings are generically expected for instance in string theories \cite{Damour:1995pd}.  In this wider class of theories, the ambiguity in the determination of the inflationary scale becomes even worse, because the set of relevant dimensionless ratios (or conformal frames) widens. Although  couplings between the inflaton and matter are useful invoked  to guarantee a successful reheating stage \cite{Kofman:1997yn}, we should also point out that they are not strictly necessary \cite{Peebles:1998qn}.

\subsection{Einstein Frame}
During inflation, the contributions of matter are negligible, so one does not need to consider  the matter Lagrangian. In that case, the action (\ref{eq:S E}) describes nothing but a conventional and canonical inflationary model. As a result, the predictions for the tensor and scalar spectra remain unaltered, and so does the standard relation  between the tensor amplitude and the scale of inflation. In particular, the tensor amplitude and the Hubble scale during inflation are related by
\begin{equation}\label{eq:A T}
	A_T=\frac{H^E_I}{M},
\end{equation}
where the script ``E" indicates Einstein-frame quantities (as we discuss later on, the tensor amplitde $A_T$ is conformal frame independent.)  Therefore, if we determine the tensor amplitude through polarization measurements \cite{Kamionkowski:1996ks,Zaldarriaga:1996xe}, we can solve for the scale of inflation in the Einstein frame,  $H^E_I = A_T M.$ Since the Hubble parameter actually changes during inflation, by $H_I$ we  mean  the value of the Hubble constant at any particular time during inflation, say, when a particular pivot scale left the horizon. 

In order to assign a numerical value to $H^E_I$ we hence need to know what $M$ is. Let us assume that after a successful reheating stage the inflaton settles down at the minimum of its potential, and that its mass at that point is sufficiently large in order for the field to remain essentially frozen at that minimum, $\varphi_*$. Then, the function $F$ takes the value $F_*\equiv F(\varphi_*)$, which we can also assume to be constant. In what follows, a star will hence label quantities after the end of inflation. Because the scalar $\varphi$ is heavy, it does not contribute to any long-ranged fifth forces. Hence,    the conventional constraints on the analogous massless  scalar-tensor theories do not apply \cite{Will:2005va}.  

In the Einstein frame, the presence of the factor $F^2_* g_{\mu\nu}$ in the matter Lagrangian of equation (\ref{eq:S E}) after the end of inflation amounts to a conformal transformation, which after appropriate field redefinitions has the only effect of scaling the mass $m$  in the Lagrangian by a factor of  $F_*$. It is thus $m_p=F_* m$ what we would call the proton mass today, and it is  this product the one  that equals about  1 GeV.\footnote{We thank Paolo Creminelli for pointing out a flaw in this part of the argument in a previous version of this manuscript.} If we now expand the metric around flat spacetime,  with canonically normalized gravitons $h_{\mu\nu}$,
\begin{equation}
	g_{\mu\nu}=\eta_{\mu\nu}+\frac{h_{\mu\nu}}{M},
\end{equation}
gravitational interactions   between non-relativistic massive particles in the Einstein frame become proportional to $m^2 F_*^2/M^2=(m_p/M)^2$. Therefore, it is $M_P\equiv M$ what we would call the Planck mass in the Einstein frame, the quantity we  measure to be around $10^{18}$~GeV today. To summarize: As conventionally argued, in the Einstein frame, equation (\ref{eq:scale}) returns the correct result.

Leaving the inflaton sector aside, the class of theories that we are studying contains two relevant mass scales: the constant Planck mass $M$, and the time-dependent proton mass $m_p=F m$. Because of the structure of the matter Lagrangian, any other scale can be expressed in terms of these two by  measuring the present values of appropriate equal-time dimensionless ratios. Say, if the theory also contains an electron, we can determine its mass $m_e$ at any time because  $m_e/m_p$ is a constant that can be measured today. Therefore, to characterize the scale of inflation during inflation we also need to calculate the ratio of the Hubble scale to the mass of the proton at the same time.  This calculation is relevant, say,  to determine whether the proton  is light or  heavy field during inflation. Because in the Einstein frame the proton mass during inflation is $m_p^I=F_I m$, along the same lines as before we readily arrive at
\begin{equation}
	H_I^E=A_T  \frac{M_P}{m_p}\left(\frac{F_*}{F_I}\right) m_p^I.
\end{equation}
The ratio $m_p/M_P\approx 10^{-18}$ is nothing but today's measurable strength of the gravitational interaction in units of the proton mass, and $A_T$ is again the observed tensor amplitude.  Since the value of the ratio $F/F_*$ is unknown,  we cannot calculate the scale of inflation in units of the proton mass at that time, even when we know the tensor amplitude.  In other words, we cannot tell whether we should regard the proton as light or heavy, or whether the Hubble scale is close to the QCD scale during inflation. In that sense, we would claim that a determination of the tensor amplitude does not completely characterize the scale of inflation. The reason is basically the presence of the new, time-dependent,  dimensionless function $F$. In a theory in which the inflaton does not determine the matter couplings $F$ remains constant, and the ambiguities we discuss here do not arise.

\subsection{Jordan Frame}

Things  look quite different in the Jordan frame, although the implications are the same. The Jordan frame is defined by the condition that the mass of the proton be independent of the inflaton. To reach this frame we need to apply the conformal transformation 
\begin{equation}\label{eq:conformal}
	g^J_{\mu\nu}= F^2 g_{\mu\nu} 
\end{equation} 
to the Einstein frame action (\ref{eq:S E}).  We then obtain a formulation of the theory in which matter couples to the Jordan frame metric $g^J_{\mu\nu}$ as desired, 
\begin{equation}\label{eq:S J}
	S_J=\int d^4x \sqrt{-g^J} \left[\frac{M^2}{2F^2}R^J+\cdots\right]
	+S_m[g^J_{\mu\nu},\psi],
\end{equation}
where we have skipped some of the  inflaton terms, which do not play much of a role here. 

By construction, in the Jordan frame the parameter $m$  defines the mass of the proton, $m_p=m\equiv1~$GeV, and it is $M_P\equiv M/F_*=10^{18}$~GeV the scale that determines what we could call today's Planck mass.  The present strength of the gravitational interactions between protons thus remains the same in both the  Einstein and the Jordan frame, because
\begin{equation}
	\frac{m_p}{M_P}=\frac{F_* m}{M}=\frac{m}{M/F_*}.
\end{equation}
In both cases we are taking ratios of the proton mass to the Planck mass, but the interpretation of these quantities in terms of the parameters contained in the Einstein or Jordan frame action changes.

Nevertheless, when dimensionless ratios of quantities evaluated at different times are involved, these  ratios take different values in different conformal frames. In a way, the calculation of these unequal-time dimensionless ratios is forced upon us, because we want to compare a mass scale during inflation to a mass scale defined today.  Suppose we decide to calculate for instance the scale of inflation in the Jordan frame in terms of the  Planck mass measured today. For  convenience, we \emph{define} this Hubble scale to be
\begin{equation}
H^J\equiv \frac{1}{a_J}\frac{da_J}{dt_J}-\frac{M}{F}\frac{d(F/M)}{dt_J}.
\end{equation}
This combination returns the Hubble scale in the Einstein frame upon the appropriate identification of times, scale factors and Planck mass parameters. If the function $F$ is nearly constant during inflation, the difference between what we call $H^J$ and the actual Hubble scale during inflation is also irrelevant. By comparing the two conformal metrics in equation (\ref{eq:conformal}) we find that  the inflationary scales and the tensor amplitudes  in both frames are related by \cite{Mukhanov:1990me,Chiba:2013mha}
\begin{equation}
	H^J=\frac{H^E}{F}, \quad
	A_T\equiv A_T^J=A_T^E.
\end{equation}
Therefore, $H$  scales as any other mass  under conformal transformations, and the tensor amplitude remains invariant as promised.  It also  follows that in terms of  today's Planck mass the scale of inflation in the Jordan frame becomes
\begin{equation}
	H_I^J=A_T \frac{F_*}{F_I} M_P.
\end{equation}
Since we do not know what the ratio $F_I/F_*$ is we cannot calculate the Jordan scale of inflation in terms of today's Planck mass. In other words, if we substitute the measured value of $M_P$ in equation (\ref{eq:scale}) we obtain the wrong Jordan frame inflationary scale. Note that the factor that quantifies the uncertainty in our determination of the Hubble scale in the Jordan frame, $F_I/F_*$, is the same as the one that enters the unknown ratio between Hubble scale and proton mass in the Einstein frame.

\section{Conclusions}

We have  argued that a measurement of the tensor amplitude does not provide us in general with a unique characterization of the scale of inflation. Although such a measurement  does  determine the scale of inflation in units  of the Planck mass at that time, it does not say anything about the scale of inflation in units of, say,  the proton mass back then. Even though  these equal-time ratios are conformal frame independent, dimensionless ratios of scales evaluated at different times do happen to depend on the conformal frame. The evaluation of these ratios is in a way forced upon us because we would like to express the scale of inflation in units of quantities we can measure today.  When we evaluate these unequal time ratios we obtain relations that only hold in particular conformal frames.  We can determine the scale of inflation in the Einstein frame in terms of the agreed value of the proton or Planck mass, but we cannot determine the same scale in  the Jordan frame in terms of the same units. Since the choice of conformal frame is just a matter of convenience,  this uncertainty is also a sign of our incomplete characterization of the inflationary scale.  While the measurement of the tensor amplitude conveys information about how relevant quantum gravity effects may be, it does not contain much  information  about eventually relevant quantum effects in the matter sector during inflation. This uncertainty can be traced back to the unknown form of the dimensionless function $F$. In theories in which the inflaton does not couple to the matter sector in the Einstein frame,  $F$ is a constant, and the obstacles we encountered here do not arise. 
\pagebreak

\noindent\emph{Note Added:} While the author was finalizing the preparation of this manuscript, Antoniadis and Patil submitted to the arxiv a preprint  that discusses  similar ideas \cite{Antoniadis:2015txa} (see their discussion between equations  (19) and (24).) Note, however, that the issues that we address here are different from the main mechanism originally proposed  in  \cite{Antoniadis:2014xva} and further discussed in \cite{Antoniadis:2015txa} .

\begin{acknowledgments}
It is a pleasure to thank Scott Watson for drawing my attention to the topic discussed in this note.  In addition, I am very grateful for useful conversations with him and Kuver Sinha.  I am also indebted to Paolo Creminelli for pointing out a flaw that prompted a substantial revision of an earlier version of this manuscript. 

\end{acknowledgments}

\end{document}